\documentclass[iop]{emulateapj}

\def\rmd{{\rm d}}

\newcommand{\oahu}{{\sc oahu}}

\shorttitle{Relativistic Hydrodynamics with Wavelets}
\shortauthors{DeBuhr et al.}

\usepackage{natbib}
\usepackage[usenames,dvipsnames]{color}

\usepackage{amsmath,amssymb}
\usepackage{graphicx}
\usepackage{tikz}
\usepackage{IEEEtrantools}

\DeclareMathOperator{\sgn}{sgn}
\DeclareMathOperator{\minmod}{minmod}

\begin{document}

\title{Relativistic Hydrodynamics with Wavelets}
\author{Jackson DeBuhr, Bo Zhang, Matthew Anderson}
\affil{Center for Research in Extreme Scale Technologies,
       School of Informatics and Computing, Indiana University, Bloomington,
       IN 47404}
\email{\{jdebuhr,zhang416,andersmw\}@indiana.edu}

\and

\author{David Neilsen, Eric W. Hirschmann}
\affil{Department of Physics and Astronomy, Brigham Young University,
       Provo, UT 84602}
\email{\{david.neilsen,ehirsch\}@physics.byu.edu}

\begin{abstract}
Methods to solve the relativistic hydrodynamic equations are a key
computational kernel in a large number of astrophysics simulations and are
crucial to understanding the electromagnetic signals that originate from
the merger of astrophysical compact objects.  Because of the many physical
length scales present when simulating such mergers,
these methods must be highly
adaptive and capable of automatically resolving numerous localized
features and instabilities that emerge throughout the computational domain
across many temporal scales.
While this has been historically accomplished with adaptive mesh refinement
(AMR) based methods, alternatives based on wavelet bases and the
wavelet transformation have recently achieved significant success in adaptive
representation for advanced engineering applications.
This work presents a new method for the integration of the
relativistic hydrodynamic equations using iterated interpolating wavelets
and introduces a highly adaptive implementation for multidimensional
simulation.  The wavelet coefficients provide a direct measure of
the local approximation error for the solution and place collocation
points that naturally adapt to the fluid flow while providing good
conservation of fluid quantities.  The resulting implementation, \oahu,
is applied to a series of demanding one- and two-dimensional problems
which explore high Lorentz factor outflows and the formation of several
instabilities, including the Kelvin-Helmholtz instability and the
Rayleigh-Taylor instability.
\end{abstract}


%-----------------------------------------------------------------------
%
%
%
%-----------------------------------------------------------------------
\section{Introduction}
\label{sec:intro}

Relativistic fluids are used to model a variety of systems in high-energy
astrophysics, such as neutron stars, accretion onto compact objects,
supernovae, and gamma-ray burst outflows.
Consequently, methods to solve the relativistic hydrodynamic
equations are a key scientific kernel in a large number
of astrophysics simulations and toolkits.
Because of the many physical length scales present when simulating
astrophysical phenomena,
these methods must also be highly
adaptive and capable of automatically resolving many localized
emerging features and instabilities throughout the computational domain
across many temporal scales.
For Eulerian fluid methods, this has been historically accomplished with
adaptive mesh refinement (AMR) based methods~\citep{BergerOliger, Andersonetal}.
Other approaches include smoothed particle hydrodynamics~\citep{Rosswog2010},
and solving the
fluid equations on a moving Voronoi mesh~\citep{Springel2010,Duffell:2011bc}.
Alternative methods based on wavelets and the
wavelet transformation have recently achieved significant success in adaptive
representation for advanced engineering applications
(\citealt{Paolucci2}, \citealt{Paolucci1}).  This has inspired and encouraged
their investigation and possible application in
relativistic hydrodynamics.
This work presents a new method for the integration of the
relativistic hydrodynamic equations using iterated interpolating wavelets
and introduces a highly adaptive implementation
for multidimensional simulations called \oahu.

The merger of two neutron stars or a black hole and a neutron star is an
astrophysical system that has attracted significant interest. The orbital
motion of the compact objects generates gravitational waves that are likely to be
observed in the new Advanced-LIGO class of gravitational wave detectors.
When operating at design sensitivity, these detectors are expected to make
many detections
of gravitational wave events each year~\citep{2010CQGra..27q3001A}.
These binaries are also expected to be
sources of significant electromagnetic emission, such as magnetosphere
interactions that give a precursor signal to merger~\citep{Palenzuela:2013hu},
kilonova events from
r-process reactions on the neutron-rich ejecta~\citep{Li:1998bw},
and short gamma-ray bursts
(SGRBs)~\citep{Berger:2013jza}.
The combination of gravitational wave and electromagnetic observations, known
as multi-messenger astronomy, should open new insights into some important
questions, such as fundamental tests of general relativity, the neutron star
equation of state, the earliest stages of supernova explosions, and
models for GRB progenitors.

Computational models of neutron star binary mergers, when compared with
observational data, are giving additional insights into such systems.
For example, a black hole-neutron star merger can produce enough ejecta
to power an SGRB, but only if the black hole has a small mass or a high
spin~\citep{Chawla:2010sw,Foucart:2013psa}.
 Moreover, simulations of neutron star binaries with a soft equation
of state produce sufficient ejecta to power an SGRB through accretion, while
those with a stiff equation of state produce much less
ejecta~\citep{2013PhRvD..87b4001H,2015arXiv150206660S,Palenzuela:2015dqa}.
Given this evidence from computer simulations,
it appears that binary neutron star
mergers and a softer nuclear equation of state may be preferred for the
production of SGRBs.
However, this continues to be an active area of research, and we
expect these results to be refined as more results become available.

Computer models of neutron star binaries can be challenging to perform.
On one hand, these models require a considerable amount of sophisticated physics to be
realistic. Such models should include full general relativity
for the dynamic gravitational field, a relativistic fluid model that
includes a magnetic field (e.g. ideal or resistive magnetohydrodynamics),
a finite-temperature
equation of state for the nuclear matter, and a radiation hydrodynamics
scheme for neutrinos. All of these components must be robust and work for
a large range of energies. A second challenge, alluded to above,
is the large range of scales that must be
resolved. The neutron star radius sets one scale, $10$--$15$~km,
as the star must be well resolved on the computational domain. Other length
scales are set by the orbital radius, $50$--$100$~km, and the gravitational
wave zone, approximately $100$--$1,000$~km. Furthermore, some fluid
instabilities can significantly increase the magnetic field strength in the
post-merger remnant, making resolutions on the scale of meters
advantageous~\citep{Kiuchi:2014hja}.
Finally, the computer models need to run efficiently on modern high performance
computers, requiring them to be highly parallelizable and scalable to run on
thousands of computational cores.

We are developing the \oahu\  code to address the challenge of simulating binary
mergers with neutron stars. A key component of this code is that
we combine a robust high-resolution shock-capturing method with
an unstructured dyadic grid of collocation points that conforms to the
features of the solution. This grid adaptivity is realized by expanding
functions in a wavelet basis and adding refinement only where the
solution has small-scale features.

Wavelets allow one to represent a function in terms of a set of basis
functions
which are localized both spatially and with respect to scale. In comparison,
spectral bases
are infinitely differentiable, but have global support; basis functions used in
finite difference or finite element methods have small compact support, but poor
continuity properties. Wavelets with compact support have been applied to the
solutions of elliptic, parabolic, and hyperbolic PDEs~\citep{Beylkin1992,
  Beylkin1998, Alpert2002, Qian1993, Qian1993a, Latto1990, Glowinski1989,
  Holmstrom1999, Dahmen1997, Urban2009, Alam2006, Chegini2011}. Wavelets have
also been applied to the solutions of integral equations~\citep{Alpert1993}.
We note that when applied to nonlinear equations, some of these previous
methods
will map the space of wavelet coefficients onto the physical space and there
compute the nonlinear terms.  They then project that result back
to the wavelet coefficients space using analytical quadrature or numerical
integration.  Our approach is rather to combine collocation methods with
wavelets thus allowing us to operate
in a single space~\citep{Bertoluzza1996, Vasilyev2000, Regele2009,
  Vasilyev1995,Vasilyev1996, Vasilyev1997}.

In astronomy, wavelets have seen extensive use in analysis tasks, from
classifying transients~\citep{Powell2015, Varughese2015}, to image
processing~\citep{Mertens2015}, and to finding solutions to nonlinear initial
value problems~\citep{Nasab2015}. They have not, however, seen much use in
solutions of PDEs in astrophysics.

This paper reports on an initial version of \oahu\  that implements the first
two elements above, concentrating on the initial tests of the fluid equations
and adaptive wavelet grid. A discussion of the Einstein equation solver and
parallelization will be presented in subsequent papers.
The organization of this paper is as follows. In section \ref{sec:methods}
we describe our model system and the numerical methods used.  Section
\ref{sec:1dtests} presents
one dimensional tests of the resulting scheme. In section \ref{sec:kelhel} we
present the results of applying the method to the relativistic
Kelvin-Helmholtz instability. Section \ref{sec:outflow} presents a stringent
test of our method as applied to a relativistic outflow that develops
Rayleigh-Taylor
generated turbulence. Finally, in section \ref{sec:summary} we summarize
results and make note of future work suggested by the method.

\section{Methods}
\label{sec:methods}

This section describes some of the numerical approaches and algorithms
used in \oahu.

\subsection{Relativistic Hydrodynamics}
\label{sec:met:rhd}

In general relativity the spacetime geometry is described by a metric
tensor $g_{\mu\nu}$, and we write the line element in ADM form as
\begin{align}
\rmd s^2 &= g_{\mu\nu}\,\rmd x^\mu \rmd x^\nu\\
         &= (-\alpha^2 + \beta_i\beta^i)\,\rmd t^2 + 2\beta_i \,\rmd t \rmd x^i
            + \gamma_{ij}\,\rmd x^i \rmd x^j.
\end{align}
We write these equations in units where the
speed of light is set to unity, $c=1$. Repeated Greek indices sum over all
spacetime coordinates, $0, 1, 2, 3$, and repeated Latin indices sum over
the spatial coordinates, $1, 2, 3$.
$\alpha$ and $\beta^i$ are functions that specify the coordinates
and $\gamma_{ij}$ is the 3-metric on spacelike hypersurfaces.
While our code is written for a completely
generic spacetime, the tests presented in this paper are all performed
in flat spacetime.  To simplify the presentation, we write the equations
in special relativity (i.e. flat spacetime) in general curvilinear
coordinates, and we set $\alpha=1$ and $\beta^i=0$.
In Cartesian coordinates, the flat space metric is the
identity $\gamma_{ij} = \mathrm{diag} (1,1,1)$, but $\gamma_{ij}$ is generally a
function of the curvilinear coordinates.

A perfect fluid in special relativity is described by a
stress-energy tensor of the form
\begin{equation}
T_{\mu\nu} = h u_\mu u_\nu + P g_{\mu\nu}\,,
\end{equation}
where $h$ is the total enthalpy of the fluid
\begin{equation}
h =  \rho (1+\varepsilon) + P.
\end{equation}
The fluid variables $\rho$, $\varepsilon$, $u^\mu$, and $P$ are the
rest mass density, the specific internal energy, the four-velocity and
the pressure of the fluid, respectively.
Once an equation of state of the form $P=P(\rho,\varepsilon)$
is adopted, the equations determining the
matter dynamics are obtained from the conservation law
$\nabla_\mu T^\mu{}_\nu = 0$ and the conservation of baryons
$\nabla_\mu(\rho u^\mu) = 0$.

We introduce the three-velocity of the fluid, $v^i$, and the Lorentz factor
$W$ by writing the four-velocity as
\begin{equation}
u^\mu = (W, Wv^i)^T.
\end{equation}
The four-velocity has a fixed magnitude $u_\mu u^\mu = -1$, which gives the
familiar relation between $W$ and $v^i$
\begin{equation}
W^2 = \frac{1}{1-v_iv^i}.
\end{equation}
We introduce a set of {\it conservative} variables
\begin{align}
\label{eq:cons_on_prim1}
\tilde D&\equiv \sqrt{\gamma}\rho W\\
\tilde S_i&\equiv \sqrt{\gamma}\left(h W^2 v_i\right)\\
\label{eq:cons_on_prim3}
\tilde \tau &\equiv \sqrt{\gamma}\left( h W^2 - P - \rho W\right).
\end{align}
These quantities correspond in the Newtonian limit to the rest mass
density, the momentum, and the kinetic energy of the fluid, respectively.
A tilde ($\tilde{\ }$) indicates that each quantity has been \textit{densitized}
by the geometric factor $\sqrt{\gamma}$, where $\gamma = \det \gamma_{ij}$.
In terms of these fluid variables, the relativistic fluid equations are
\begin{align}
&\partial_t \tilde D + \partial_i \left( \tilde D v^i \right) = 0 \\
&\partial_t \tilde S_j
   + \partial_i\left(v^i \tilde S_j + \sqrt{\gamma} P \gamma^i{}_j\right)
   = {}^{3}\Gamma^i{}_{kj}\left( v^k \tilde S_i
                            + \sqrt{\gamma} P \gamma^{k}{}_i\right)
     \\
&\partial_t \tilde \tau + \partial_i \left( \tilde S^i -  \tilde D v^i \right)
 =0,
\end{align}
where ${}^{3}\Gamma^i{}_{jk}$ are the Christoffel symbols associated with the
spatial metric $\gamma_{ij}$.

The fluid equations of motion can be written in balance law form
\begin{equation}
\partial_t {\bf u} + \partial_i{\bf f}^i\left({\bf u}\right) = {\bf s}({\bf u}),
\end{equation}
where $\mathbf{u}$ is the state vector of conserved variables
and ${\bf f}^i$ are the fluxes
\begin{equation}
\mathbf{u} = \left(\begin{array}{c}
     \tilde D\\
     \tilde S_i\\
     \tilde \tau
     \end{array}
 \right), \qquad
\mathbf{f}^i = \left(\begin{array}{c}
     \tilde D v^i\\
     v^i\tilde S_j + \sqrt{\gamma} P \gamma^i{}_j\\
     \tilde S^i - \tilde D v^i
     \end{array}
 \right).
\end{equation}
The fluid equations in curvilinear coordinates have geometric source terms,
which are included in $\bf s$.

Finally, the system of equations is closed with an equation of state.
We use the $\Gamma$-law equation of state
\begin{equation}
P = (\Gamma -1)\rho\,\varepsilon,
\end{equation}
where $\Gamma$ is the adiabatic constant.

\subsection{Sparse Field Representation}
\label{sec:met:sfr}

In this section we describe how we construct the sparse, adaptive representation
of fields. The essential ingredients are the iterative interpolation of
\cite{Deslauriers1989} and the wavelet representation of \cite{Donoho1992}.
This presentation follows that in \cite{Holmstrom1999}. We begin with the
one dimensional case; see Section~\ref{sec:met:2d3d} for the generalization to
higher dimensions.

\begin{figure}
\begin{center}
\includegraphics{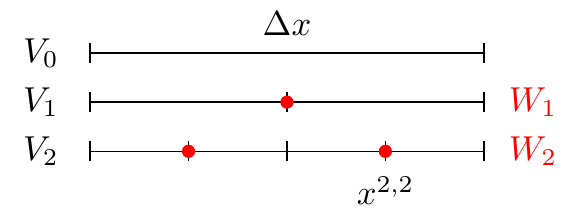}
\end{center}
\caption{The one dimensional nested dyadic grids, $V_j$. This example has $N=1$
and shows up to level $j=2$. The point at $j=2$, $k=2$ is labeled. In red are
shown those points that are part of the alternate grids, $W_j$, which are
defined for $j > 1$.}
\label{fig:basicgrid}
\end{figure}

The method begins with a nested set of dyadic grids, $V_j$ (see
Figure~\ref{fig:basicgrid}):

\begin{equation}
V_j = \left\{x^{j,k} : x^{j,k} = 2^{-j} k \Delta x \right\}.
\end{equation}

\noindent Here $\Delta x$ is the spacing at level $j = 0$, called the base
level, and $k$ is an integer indexing the points within the various grid levels.
Notice that the points in $V_j$ will also appear in all higher level grids
$V_l$ (where $l > j$). The points of even $k$ at level $j$ will also be in the
grid at level $j-1$. If the overall domain size is $L$, and there are $N + 1$
points in grid $V_0$, then $\Delta x = L / N$.

Starting with a set of field values at level $j$, $u^{j,k}$, we can extend
these values to higher levels of the grid using interpolation. For those
points in $V_{j+1}$ that are also in $V_j$, we just copy the value from the
coarser grid: $u^{j+1,2k} = u^{j,k}$. The previous is also the means by which
the field values can be restricted to coarser levels: points at coarser
levels have values copied from finer levels. For the points first appearing in
grid $V_{j+1}$ we take the nearest $p$ field values from grid $V_j$ and
interpolate:

\begin{equation}
u^{j+1,2k+1} = \sum_m h^{j+1,2k+1}_{j,m} u^{j,m}.
\end{equation}

\noindent Here $h^{l,m}_{j,k}$ are the coefficients for interpolation from
level $j$ to level $l$. In practice, for a given $k$, only a small number of
these coefficients are nonzero. In this work we use $p = 4$, so we have

\begin{align}
u^{j+1,2k+1} = & -\frac{1}{16} u^{j,k-1} + \frac{9}{16} u^{j,k} \nonumber\\
 & + \frac{9}{16} u^{j,k+1} - \frac{1}{16} u^{j,k+2}.
\end{align}

\noindent The previous applies in the interior of the grid. Near the boundaries,
little changes, except the nearest points are no longer symmetric around the
refined point, and the coefficients in the sum are different. Having advanced
the field values to grid $V_{j+1}$, the procedure can be iterated to advance
the field to $V_{j+2}$. In this way, any level of refinement can be
achieved from the initial sequence, and when performed \emph{ad infinitum}
produces a function on the interval $[0, L]$ (see \cite{Donoho1992} for details
about the regularity of these functions).

The linearity of this procedure suggests a natural set of basis functions for
iterated interpolation functions formed from Kronecker sequences.
That is, start with a sequence that has a single $1$ at level $j$ and
interpolate, $\phi_{j,k}(x^{j,l}) = \delta^l_k$.
The resulting function, $\phi_{j,k}(x)$ has a number of properties, among
which is the two-scale relation:

\begin{equation}
\phi_{j,k}(x) = \sum_l c^l \phi_{j+1,l}(x).
\end{equation}

\begin{figure}
\includegraphics[width=7.cm,angle=0]{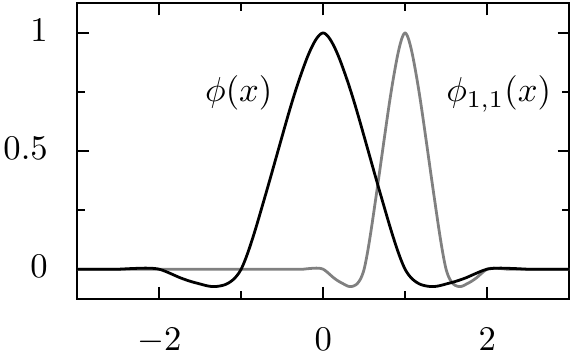}
\caption{The fundamental solution of the iterated interpolation for $p = 4$,
$\phi(x)$ (black) and a basis element in level one, $\phi_{1,1}(x)$ (gray).
Each basis function is a scaled, translated version of the fundamental
solution.}
\label{fig:basis}
\end{figure}

\noindent One step of the interpolation will produce a sequence on $V_{j+1}$,
which can be written as a weighted sum of the Kronecker sequences on $V_{j+1}$.
Indeed, the weights $c^l$ are easy to find from the interpolation. For $p = 4$
these are $c^l = \{..., 0, -1/16, 0, 9/16, 1, 9/16, 0, -1/16, 0, ...\}$.
Each of these functions is a scaled, translated version of a single
function $\phi(x)$, shown in Figure~\ref{fig:basis}, called the fundamental
solution of the interpolation:
$\phi_{j,k}(x) = \phi(2^j x / \Delta x - k)$.

The previous functions can be used to form bases for each level of the grid
separately. The two-scale relation prevents the full set from being a basis on
the full set of collocation points. In particular, note that for each location in the
grid, $x^{j,k}$ there is one such function. However, certain locations are
represented on multiple levels, for example $x^{j,k} = x^{j+m, 2^m k}$ for all
$m > 0$. To form a basis for the full grid, introduce the alternate higher
level grids, $W_j$, for $j > 0$:

\begin{equation}
W_j = \left\{ x^{j,k} : x^{j,k} = 2^{-j} k \Delta x, \, k \mbox{~odd} \right\}.
\end{equation}

\noindent The $W_j$ grids are the $V_j$ grids with the points from earlier
levels removed (see the red points in Figure~\ref{fig:basicgrid}).
The set of grids $\{V_0, W_j\}$ now has each point represented
exactly once. Forming a basis for these points is achieved by taking the
set of $\phi_{j,k}(x)$ functions that correspond to the points in the base
grid, and these higher level alternate grids, $W_j$. With this basis, we
can represent a field, $u$, as follows:

\begin{align}
u(x) = & \sum_{k \in S_0} u^{0,k} \phi_{0,k}(x) \nonumber\\
 & + \sum_{j=1} \sum_{k \in S_j} d^{j,k} \phi_{j,k}(x).
\end{align}

\noindent where $S_0 = \{0, 1, ..., N\}$ is the index set for grid $V_0$ and
$S_j = \{1, 3, ..., 2^j N - 1\}$ is the index set for grid $W_j$.

The previous is the interpolating wavelet expansion of the field. Intuitively,
the expansion contains the coarse picture (level 0) and refinements of that
picture at successively finer levels. The expansion coefficients $u^{0,k}$ are
just the field values at the base level points: $u^{0,k} = u(x^{0,k})$. We can
extend this notation to include $u^{j,k} = u(x^{j,k})$. The coefficients
$d^{j,k}$, called {\it wavelet coefficients}, are computed by comparing the
interpolation from the previous level to the field value, $u^{j,k}$.
In particular if we denote the interpolated value from level $j$ at a
level $j+1$ point, $\tilde{u}^{j+1,k} = P(x^{j+1,k}, j)$, then

\begin{equation}
d^{j, k} = u^{j, k} - P(x^{j, k}, j - 1).
\end{equation}

\noindent Intuitively, the wavelet coefficient measures the failure of the
field to be the interpolation from the previous level. The
previous is also called the forward wavelet transformation. This transformation
starts with field values on the multi-level grid, and produces wavelet
coefficients. The transformation can be easily inverted by rearranging the
equation, and computing field values given the wavelet coefficients.

There are two descriptions of the field on the multi-level grid. The first,
called the {\it Point Representation} is the set of values $\{u^{j,k}\}$. The
second, called the {\it Wavelet Representation} is the set of values
$\{u^{0,k},d^{j,k}\}$.  These representations can be made sparse via
thresholding. Starting with the Wavelet Representation, and given a threshold
$\epsilon$,
the {\it Sparse Wavelet Representation} is formed by removing those
points whose wavelet coefficient are below some threshold: $|d^{j,k}| <
\epsilon$. This amounts to discarding those points that are well approximated
by interpolation.  This naturally cuts down the number of points in the grid,
and introduces an \emph{a priori} error bound~\citep{Donoho1992} on the
representation of the field. The points whose values are kept are called
{\it essential} points. The level 0 points are always kept and are always
essential. The field values at the essential points form the {\it Sparse Point
Representation.}

\subsection{Higher Dimensions}
\label{sec:met:2d3d}

In multiple dimensions, the construction is not much more involved. The basis
functions are taken to be the products of the 1-dimensional functions:

\begin{equation}
\phi_{j,\vec{k}}(x, y, z) = \phi_{j,k_x}(x) \phi_{j,k_y}(y) \phi_{j,k_z}(z).
\end{equation}

\begin{figure}
\begin{center}
\includegraphics{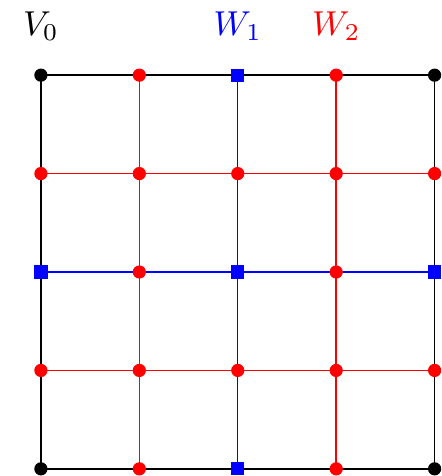}
\end{center}
\caption{A slice of the 3d grid. The black points (circles) are those in the
  base grid, the blue points (squares) belong to level 1, and the red points
  (circles) belong to level 2. Notice that some points at level 1 line up with
  points at level 0, and that some points at level 2 line up with points at
  level 1.}
\label{fig:grid2d}
\end{figure}

\begin{figure}
\begin{center}
\includegraphics{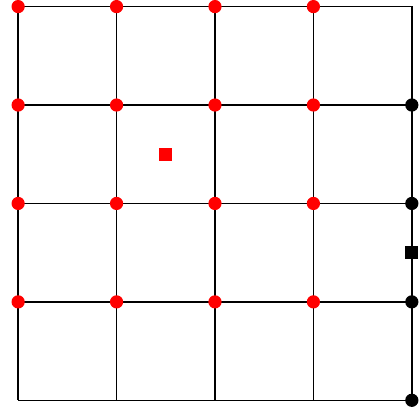}
\end{center}
\caption{A portion of a slice of the 3d grid at level $j$.
 Shown are those points at level
$j$ (circles) that contribute to the wavelet transformation for the marked
points at level $j+1$ (squares). Depending on the relative placement of the
level $j+1$ points to the level $j$ grid, the wavelet transformation will use
$p$ (black) or $p^2$ (red) points. In three dimensions this extends to cases
that also use $p^3$ points.}
\label{fig:grid2dwavelet}
\end{figure}

\noindent In the previous, $\vec{k} = (k_x, k_y, k_z)$ is the set of three
indices required to label a three dimensional grid (see Figure~\ref{fig:grid2d}).
Another way to construct these functions is by interpolation
in multiple dimensions. Depending on the location of the level $j+1$ point in
the level $j$ grid, this interpolation will involve $p, p^2$ or $p^3$ terms
(Figure~\ref{fig:grid2dwavelet}).  The wavelet coefficient for a point is again
computed as the difference between the field value and the interpolated value,
it is just that the interpolation often contains more terms.

The rest of the method goes through as one might expect. The field is expanded
in terms of these basis functions:

\begin{align}
u(x, y, z) = & \sum_{\vec{k}} u^{0,\vec{k}} \phi_{0,\vec{k}}(x, y, z) \nonumber\\
 & + \sum_{j=1} \sum_{\vec{k}} d^{j,\vec{k}} \phi_{j,\vec{k}}(x,y,z).
\end{align}

\noindent The sparse representation is formed by removing those points whose
wavelet coefficient have a magnitude less than the prescribed error threshold,
$\epsilon$.

\subsection{Conservation}
\label{sec:met:cons}

It is possible to measure the conservation of the fields being evolved
using the wavelet basis. In particular, the quantity

\begin{equation}
\int_{\Sigma} u(x) dx,
\end{equation}

\noindent where $\Sigma$ is the computational domain, is straightforward to
compute using the standard expansion of the field in the wavelet basis. Making
the substitution yields

\begin{align}
\int_{\Sigma} u(x) dx = & \sum_k u^{0,k} \int_D \phi_{0,k}(x) dx\nonumber\\
 & + \sum_{j=1} \sum_{k \in S_j} d^{j,k} \int_D \phi_{j,k}(x) dx.
\end{align}

\noindent Given the integrals of the basis functions, we can easily compute
the total amount of some quantity.

That each basis element is a scaled, translated version of the fundamental
solution to the interpolation implies that, ignoring edges of the computational
domain for the moment,

\begin{equation}
\int \phi_{j,k}(x) dx = \frac{\Delta x}{2^j} \int \phi(x) dx.
\end{equation}

\noindent The prefactor takes into account the difference in spacing for the
two functions. The fundamental solution is defined for a spacing of 1.
It is straightforward to show that the integral of the fundamental solution is
1 by defining a sequence of approximations to the integral using Riemann sums:

\begin{equation}
I_{(j)} = \sum_k \frac{1}{2^j} \phi(x^{j,k}).
\end{equation}

\noindent The interpolation property of the fundamental solution makes it
easy to show that $I_{(j)} = I_{(j+1)}$. Then, given the starting point that
$I_{(0)} = 1$ we have that

\begin{equation}
\int \phi(x) dx = \lim_{j \rightarrow \infty} I_{(j)} = 1.
\end{equation}

\noindent Thus,

\begin{equation}
\int \phi_{j,k}(x) dx = \frac{\Delta x}{2^j}.
\end{equation}

We can use this in the expansion of the field to write

\begin{equation}
\int_{\Sigma} u(x) dx = \sum_k u^{0,k} \Delta x  + \sum_{j=1} \sum_{k \in S_j} d^{j,k} \frac{\Delta x}{2^j}.
\end{equation}

\noindent This expression allows the monitoring of the conserved quantities
during the simulation. In every case examined, the conservation is good to
the level of the chosen $\epsilon$.

Near the edges of the computational grid, the basis elements no longer have
unit integrals. To compute these integrals, it is necessary to make use of the
two-scale relation for the basis to compute partial integrals:

\begin{equation}
I_a^b \equiv \int_a^b \phi(x) dx
\end{equation}

\noindent where $a, b$ are integers, one of which might be infinite. Some are
simple, $I_0^{\infty} = 0.5$ due to the symmetry of the basis, but others
require setting up a linear system using the two-scale relation. Once the set
of partial integrals is computed, the basis elements near the edges are written
as the sum of an extended basis that stretches past the edge of the
computational domain. It is an exercise in algebra to show that each of the
original basis elements modified by the edges can be written as a sum of
these extended basis elements. The extended basis elements are again
translated,
scaled versions of the fundamental solution, so we can use the partial integrals
for only the interval inside the original domain to compute the integral of the
original basis elements. For the case of $p=4$, with superscripts
labeling the location in the grid, we find that

\begin{align}
I_j^0 = \frac{121}{360} \frac{\Delta x}{2^j} & \mbox{~~} & I_j^1 = \frac{462}{360} \frac{\Delta x}{2^j} \nonumber\\
I_j^2 = \frac{303}{360} \frac{\Delta x}{2^j} & \mbox{~~} & I_j^3 = \frac{374}{360} \frac{\Delta x}{2^j}.
\end{align}

\noindent Similar expressions hold at the largest $k$ values.

In multiple dimensions, because the basis functions are simple products,
the integral of the basis is just the product of the integrals in each
direction.

\subsection{Fluid Methods}
\label{sec:met:fluids}

As mentioned in the introduction, one goal of this project is to develop
a fully relativistic fluid code to study the binary mergers of compact objects.
This involves solving both the Einstein equations of general relativity for
the spacetime geometry and coordinate conditions, and the relativistic
fluid equations for an arbitrary geometry. With neutron star binary mergers
as our model problem, we choose a numerical algorithm that satisfies
the following conditions:
(1) The fluid density and pressure in a neutron star
spans several orders of magnitude, and ejecta from the stars can reach
speeds near the speed of light, so we choose a robust high-resolution
shock-capturing method.
(2) The proper calculation of normal modes for neutron stars requires
high-order reconstruction methods for the fluid variables so
we implement both PPM and MP5 reconstructions.
(3) The calculation of characteristic variables is
computationally intensive for relativistic fluids, especially for relativistic
MHD, so we choose a central scheme with an approximate Riemann solver.
(4) The Einstein equations---fundamentally equations for classical fields
often written in a second-order formulation---are most naturally
discretized using finite differences so we choose a finite-difference fluid
method to simplify coupling the two sets of equations.

In this section we assume a uniform grid of points labeled with an
index, i.e., $x_j = x_{\rm min} + j\Delta x$, and a function evaluated
at point $x_j$ has the shorthand notation $f_j = f(x_j)$.

Our numerical method for solving the fluid equations
is based on the finite-difference Convex ENO (CENO)
scheme of Liu and Osher for conservation laws~\citep{LiuOsher}.
For simplicity we present the
method for a one-dimensional problem. The extension to multiple dimensions
is done by differencing each dimension in turn, as
discussed in Section~\ref{sec:time_integration}.
We write the conservation law in semi-discrete form as
\begin{equation}
\frac{\rmd \mathbf{u}_i}{\rmd t} =
     - \frac{\Delta t}{\Delta x}
        \left(\hat{\mathbf{f}}_{i+1/2} - \hat{\mathbf{f}}_{i-1/2} \right),
\end{equation}
where $\hat{\mathbf{f}}_{i+1/2}$ is a consistent numerical flux function
\begin{align}
\hat{\mathbf{f}}_{i+1/2}
              &= \hat{\mathbf{f}}(\mathbf{u}_{i-k}, \ldots,  \mathbf{u}_{i+m})\\
      \mathbf{f}(\mathbf{u})
              &= \hat{\mathbf{f}}(\mathbf{u}, \ldots ,\mathbf{u}).
\end{align}
Liu and Osher base the CENO method on the local Lax-Friedrichs (LLF)
approximate Riemann solver, and they use a ENO interpolation scheme
to calculate the numerical flux functions $\mathbf{f}_{i+1/2}$.  In previous
work we have found that the CENO scheme is too dissipative to reproduce
the normal modes of neutron stars~\citep{Andersonetal},
so we use the HLLE numerical flux~\citep{Harten,Einfeldt}
in place of LLF, and we use higher-order finite volume reconstruction
methods, such as PPM~\citep{Colella:1982ee} and MP5~\citep{Suresh1997}.

The numerical flux $\hat{\mathbf{f}}_{i+1/2}$ requires the fluid state at
$\mathbf{u}_{i+1/2} = \mathbf{u}(x_{i+1/2})$.  We use the fluid variables
near this point to reconstruct both left and right states at the midpoint,
$\mathbf{u}_{i+1/2}^\ell$ and $\mathbf{u}_{i+1/2}^r$, respectively. The
numerical flux can then be written in terms of these new states as
$\mathbf{f}_{i+1/2} = \mathbf{f}(\mathbf{u}_{i+1/2}^\ell,\mathbf{u}_{i+1/2}^r)$.
We have
implemented piece-wise linear (TVD) reconstruction,
the Piece-wise Parabolic Method (PPM),
and MP5 reconstruction. The MP5 reconstruction method usually gives
superior results compared to the other methods, so we have used this
reconstruction for all tests in this paper, except for Case IV below.
As we use a central scheme for the approximate Riemann solver, we
reconstruct each fluid variable separately. Moreover, given the difficulty
of calculating primitive variables ($\rho$, $v^i$, $P$) from the conserved
variables ($D$, $S_i$, $\tau$), we reconstruct the primitive variables
$(\rho, v^i, P)$,
and then calculate corresponding conserved variables.

The MP5 method is a polynomial reconstruction of the fluid state that preserves
monotonicity~\citep{Suresh1997,Moesta:2013dna}.
It preserves accuracy near extrema and is computationally
efficient. The reconstruction of a variable $q$ proceeds in two steps.
We first calculate an interpolated value for the state $q^\ell_{i+1/2}$,
called the original value. In the second step, limiters are applied to the
original value to prevent oscillations, producing the final limited value.
The original value at the midpoint is
\begin{equation}
q^\ell_{i+1/2}  = \frac{1}{60}\left(
    2q_{i-2} - 13 q_{i-1} + 47 q_{i} + 27 q_{i+1} - 3 q_{i+2}
\right).
\end{equation}
We then compute a monotonicity-preserving value
\begin{equation}
q^{MP} = u_i + \minmod(q_{i+1} - q_i, \tilde\alpha(q_i- q_{i-1})),
\end{equation}
where $\tilde\alpha$ is a constant which we set as $\tilde\alpha = 4.0$.
The minmod function gives the argument with the smallest magnitude
when both arguments have the same sign
\begin{equation}
\minmod(x,y) = \frac{1}{2}\left(\sgn\left(x\right)
       + \sgn\left(y\right)\right)\min\left(|x|,|y|\right).
\end{equation}
The limiter is not applied to the original value when
\begin{equation}
(q^\ell_{i+1/2} - q_i)(q^\ell_{i+12} - q^{MP}) \le \varpi|q|,
\label{eq:mp5_condition}
\end{equation}
where $\varpi=10^{-10}$ and $|q|$ is the $L_2$ norm of $q_i$ over the
stencil points $\{q_{i-2},\ldots, q_{i+2}\}$.  The $|q|$ factor does
not appear in the original algorithm, but we follow \citet{Moesta:2013dna}
in adding this
term to account for the wide range of scales in the different fluid
variables.

When condition Eq.~(\ref{eq:mp5_condition}) does not hold, we apply a limiter
to the original value. We then compute the second derivatives
\begin{align}
D^-_i &= q_{i-2} - 2q_{i-1} + q_{i}\\
D^0_i &= q_{i-1} - 2q_{i} + q_{i+1}\\
D^+_i &= q_i - 2q_{i+1} + q_{i+2},
\end{align}
and
\begin{align}
D^{M4}_{i+1/2} = \minmod(4D^0_i-D^+_i, 4D^+_i - D^0_i,D^0_i,D^+_i)\\
D^{M4}_{i-1/2} = \minmod(4D^0_i-D^-_i, 4D^-_i - D^0_i,D^0_i,D^-_i).
\end{align}
The minmod function is easily generalized for an arbitrary number of
arguments as
\begin{equation}
\minmod(z_1,\ldots,z_k) = s\min\left(|z_1|,\ldots,|z_k|\right),
\end{equation}
where
\begin{align}
s &= \frac{1}{2}\left(\sgn(z_1)+\sgn(z_2)\right)\times\nonumber\\
   &\quad \left|\frac{1}{2}\left(\sgn(z_1)+\sgn(z_3)\right)\times\ldots\times
    \frac{1}{2}\left(\sgn(z_1)+\sgn(z_k)\right) \right|.
\end{align}
We then compute the following quantities
\begin{align}
q^{UL} &= q_i + \alpha\left(q_i - q_{i+1}\right)\\
q^{AV} &= \frac{1}{2}\left(q_i + q_{i+1}\right)\\
q^{MD} &= q^{AV} - \frac{1}{2} D^{M4}_{i+1/2}\\
q^{LC} &= q_{i} + \frac{1}{2}\left(q_i - q_{i-1}\right)
                + \frac{4}{3} D^{M4}_{i-1/2},
\end{align}
to obtain limits for an accuracy-preserving constraint
\begin{align}
q_{\min} = \max\left(\min\left(q_i, q_{i+1}, q^{MD}\right),
                     \min\left(q_i, q^{UL}, q^{LC}\right)\right)\\
q_{\max} = \min\left(\max\left(q_i, q_{i+1}, q^{MD}\right),
                     \max\left(q_i, q^{UL}, q^{LC}\right)\right).
\end{align}
Finally, the limited value for the midpoint is
\begin{equation}
q^{\ell,{\rm Lim}}_{i+1/2}
       = q^\ell_{i+1/2} + \minmod\left(q_{\min} - q^\ell_{i+1/2},
                                       q_{\max} - q^\ell_{i+1/2}\right).
\end{equation}

To compute the right state $q^r_{i-1/2}$, we repeat the algorithm but
reflect the stencil elements about the center, replacing
$\{q_{i-2}, q_{i-1}, q_i, q_{i+1}, q_{i+2}\}$
with
$\{q_{i+2}, q_{i+1}, q_i, q_{i-1}, q_{i-2}\}$.

The HLLE approximate Riemann solver is
a central-upwind flux function that
uses the maximum characteristic speeds in each direction to calculate
a solution to the Riemann problem~\citep{Harten,Einfeldt}
\begin{align}
\hat{\mathbf{f}}_{i+1/2}^{\mathrm{HLLE}} &=
\frac{\lambda^+_r \mathbf{f}(\mathbf{u}_{i+1/2}^\ell)
    - \lambda^-_\ell\mathbf{f}(\mathbf{u}_{i+1/2}^r)}{\lambda^+_r - \lambda^-_\ell} \nonumber\\
  &+ \frac{\lambda^+_r \lambda^-_\ell(\mathbf{u}_{i+1/2}^r
              - \mathbf{u}_{i+1/2}^\ell)}{\lambda^+_r - \lambda^-_\ell},
\end{align}
where $\lambda^+$ and $\lambda^-$ represent the largest characteristic
speeds at the interface in the right and left directions, respectively.

The largest and smallest characteristic speeds of the relativistic
fluid in flat spacetime in the direction $x^i$ are
\begin{align}
\lambda_{\pm} &=
   \frac{1}{1-v^2c_s^2}\left\{ v^i(1-c_s^2)\right.\nonumber\\
 &\pm \left.\sqrt{c_s^2(1-v^2)\left[\gamma^{ii}(1-v^2 c_s^2) - v^iv^i(1-c_s^2)\right]}
        \right\},
\end{align}
where the sound speed $c_s$ is
\begin{equation}
c_s^2 = \frac{1}{h}
         \left.\left(\frac{\partial P}{\partial \rho}\right)\right|_\epsilon
       + \frac{P}{\rho^2 h}
         \left.\left(\frac{\partial P}{\partial \epsilon}\right)\right|_\rho.
\end{equation}

\subsection{Time Integration}
\label{sec:time_integration}

The conservation equations are written in semi-discrete form:
\begin{align}
\frac{\rmd \mathbf{u}_{i,j,k}}{\rmd t} =
   &- \frac{1}{\Delta x}\left(\hat{\mathbf{f}}^1{}_{i+1/2,j,k}
                            - \hat{\mathbf{f}}^1{}_{i-1/2,j,k} \right)\nonumber\\
   &- \frac{1}{\Delta y}\left(\hat{\mathbf{f}}^2{}_{i,j+1/2,k}
                            - \hat{\mathbf{f}}^2{}_{i,j-1/2,k} \right)\nonumber\\
   &- \frac{1}{\Delta z}\left(\hat{\mathbf{f}}^3{}_{i,j,k+1/2}
                            - \hat{\mathbf{f}}^3{}_{i,j,k-1/2} \right)\nonumber\\
   &+ \mathbf{s}\left( \mathbf{u_i} \right)
\end{align}
where $\hat{\mathbf{f}}_{i+1/2}$ is the numerical flux. The flux
functions in each direction are evaluated separately.
The sparse wavelet representation leads to a scheme for integrating a system
of differential equations in time by using the method of lines.
The coefficients
in the expansion become time dependent and can be integrated in time using
any standard time integrator. In this work, the classical fourth-order
Runge-Kutta method is used which has a CFL coefficient $\lambda_0 = 2/3$.
The velocity and characteristic speeds of the fluid are bounded by the speed
of light, so the
time step $\lambda = c\Delta t / \Delta x$ is bounded by the
CFL coefficient $\lambda \le \lambda_0$.

As the physical state evolves during the simulation, the set of essential
points will change. This means that the method needs to support the promotion of
a point to becoming essential and the demotion of a point from being essential.
To allow for such changes to the set of essential points, so-called
{\it neighboring points} are added to the grid.
These are those points adjacent to an essential point at the next finer
level. In a sense, these points are sentinels waiting to become essential.
Given that the points at level 0 are always essential, the points at level 1
will all be at least neighboring, and in this way, both the level 0 and level
1 grids will be fully occupied. Both
neighboring and essential points participate in time integration and, for this
reason, are called {\it active} points.

At the end of each timestep, every active point has its wavelet transformation
computed. Essential points that no longer exceed the error threshold are
demoted, and neighboring points that exceed the threshold are promoted.
Neighboring points promoted to being essential points will thus require their own
neighboring points at the next finer level. In this way, as the solution
develops features on finer scales, the grid adapts, adding
points to the grid exactly where the resolution is needed. Initially, the field
values for a neighboring point can be taken from the initial conditions.
Neighbors added after the initial time slice are given field values from the
inverse wavelet transformation. That is, they are given a wavelet coefficient
of zero, so their fields are equal to the interpolation from the previous level.

Finally, there is a third class of grid points, called {\it nonessential}
points,
that are required to fill out wavelet or other computational stencils of
essential and neighboring points. Nonessential points do not participate in
time integration and are given values via interpolation.

In practice, an upper limit on refinement, $j_{\rm max}$, must be specified. During
an evolution, it can happen that the solution naturally
attempts to refine past $j_{\rm max}$. There are two options in this case. The
first is for the code to complain and die, and the second is for the code to
complain and warn the user. Taking the former approach assures that the error
bounds implied by the chosen $\epsilon$ are not violated, because if the grid
attempts to add a point at level $j_{\rm max} + 1$ it is because the point is
needed. However, if there is a discontinuity in the solution, no level of
refinement will be sufficient to satisfy the refinement criterion. One
possible solution to this problem is to smooth over discontinuities in initial
data, and add some viscosity to prevent their formation during the simulation.
However, we are interested in sharp features that develop during the
simulation and it is for that reason that we use a fluid reconstruction
and numerical flux function so as to allow for these sharp features. Thus,
with a method that allows for discontinuities, we can never fully satisfy
the refinement criterion if a discontinuity develops, and so the grid will
want to refine forever. As a result, we take the practical step of limiting
the maximum refinement level.

\subsection{Primitive Solver}
\label{sec:met:prim}

An important aspect of any relativistic hydrodynamics code is the inversion
between
the conserved and the primitive variables.  Because the equations are
written in conservation form, the conserved variables are the evolved
variables.  However, the primitive variables are needed as part of the
calculation.  For Newtonian fluids, this inversion from the conserved to
the primitive variables is algebraic, can be done in closed form and results
in a unique solution for the primitive variables provided the conserved
variables take on physical values.  Such is not the case in relativistic
situations.  As a result, a number of related procedures can be found in
the literature to effect this
inversion~\citep{Duez2005, Etienne2012, Noble2006}.
Due to its importance, we sketch our approach to performing the
inversion.

For this discussion, we revert to the undensitized form of the conserved
variables $(D, S_i, \tau)$.  In terms of the fluid primitives
$(\rho, v^i, P)$,
and for our chosen ($\Gamma$-law) equation of state (with $1<\Gamma\le2$),
the conserved variables are given by the undensitized version of
Eqs. (\ref{eq:cons_on_prim1}-\ref{eq:cons_on_prim3}).
The
inversion can be reduced to a single equation for $x=h W^2/(\tau+D)$,
namely,
\begin{equation}
- \left( x - {\Gamma\over2} \right)^2 + {\Gamma^2 \over 4} - (\Gamma-1) \beta^2 = (\Gamma-1) \, \delta \, \sqrt{ x^2 - \beta^2 },
\end{equation}
where we have defined
\begin{equation}
\beta^2 = \frac{S^2}{(\tau+D)^2}, \qquad \delta = \frac{D}{\tau+D}.
\end{equation}
Note that the left hand side, call it $f(x)$, is a downward pointing quadratic
while the right hand side includes the square root of another quadratic,
$g(x) = x^2 - \beta^2$, which has roots $\pm|\beta|$.  Solving for $x$ amounts
to finding the intersections of $f(x)$ and $(\Gamma-1)\delta\sqrt{g(x)}$.
For a physical solution, $x>|\beta|$, the largest root of $g(x)$.
Therefore, there is a single intersection
bracketed by this root of $g(x)$ and the larger root of $f(x)$, namely
$|\beta| < x < x^{*}$ where
\begin{equation}
x^{*} = {\Gamma \over 2}
     + \left[ {\Gamma^2\over4} - (\Gamma-1) \beta^2 \right]^{1/2}.
\end{equation}
That this root of $f(x)$ is real is guaranteed by the dominant energy
condition, given
here as $\beta^2<1$.  Hence a unique solution to the primitive inversion
exists in the pure hydrodynamics case provided the conserved variables satisfy
this inequality.
Using a straightforward Newton's method allows us to solve for the primitive
variables in virtually every case.  Occasionally, when the inequality is
violated, rescaling $\beta^2$ to bring it within physical bounds is sufficient
to allow the primitive solve to proceed to a solution.

%-----------------------------------------------------------------------
%
%
%
%-----------------------------------------------------------------------
\section{One Dimensional Tests}
\label{sec:1dtests}

\cite{LoraClavijo2013} outlines a number of simple one-dimensional Riemann
problems for relativistic hydrodynamics. Each of the four cases outlined
was run with \oahu, and the results are reported below. In each case, the
base grid was chosen to have $N=10$, and we have varied both the maximum
number of refinement levels and the value of $\epsilon$. For each Riemann
problem, we
take the overall domain to be from $x = -1$ to $x = 1$ with the two states
separated at $x = 0$. The primitive fields on each side of the problem for the
four cases are given in Table~\ref{tab:cases}.

\begin{table}[htb]
  \centering
  \begin{tabular}{|lc|ccc|}
    \hline
     Case & & $v$ & $\rho$ & $p$ \\
    \hline
     I & $x < 0$ & $0$ & $10$ & $13.33$ \\
       & $x > 0$ & $0$ & $1$ & $10^{-6}$ \\
    \hline
     II & $x < 0$ & $0$ & $1$ & $10^{-6}$ \\
        & $x > 0$ & $0$ & $10$ & $13.33$ \\
    \hline
     III & $x < 0$ & $-0.2$ & $0.1$ & $0.05$ \\
         & $x > 0$ & $0.2$ & $0.1$ & $0.05$ \\
    \hline
     IV & $x < 0$ & $0.999999$ & $0.001$ & $3.333 \times 10^{-9}$ \\
        & $x > 0$ & $-0.999999$ & $0.001$ & $3.333 \times 10^{-9}$ \\
    \hline
  \end{tabular}
  \caption{Initial state for four Riemann problems tested with \oahu. The
  simulated domain is the interval $x \in [-1,1]$, and the separation
  between the left and right states is at $x=0$.}
  \label{tab:cases}
\end{table}

\begin{figure*}
  \includegraphics[width=\textwidth]{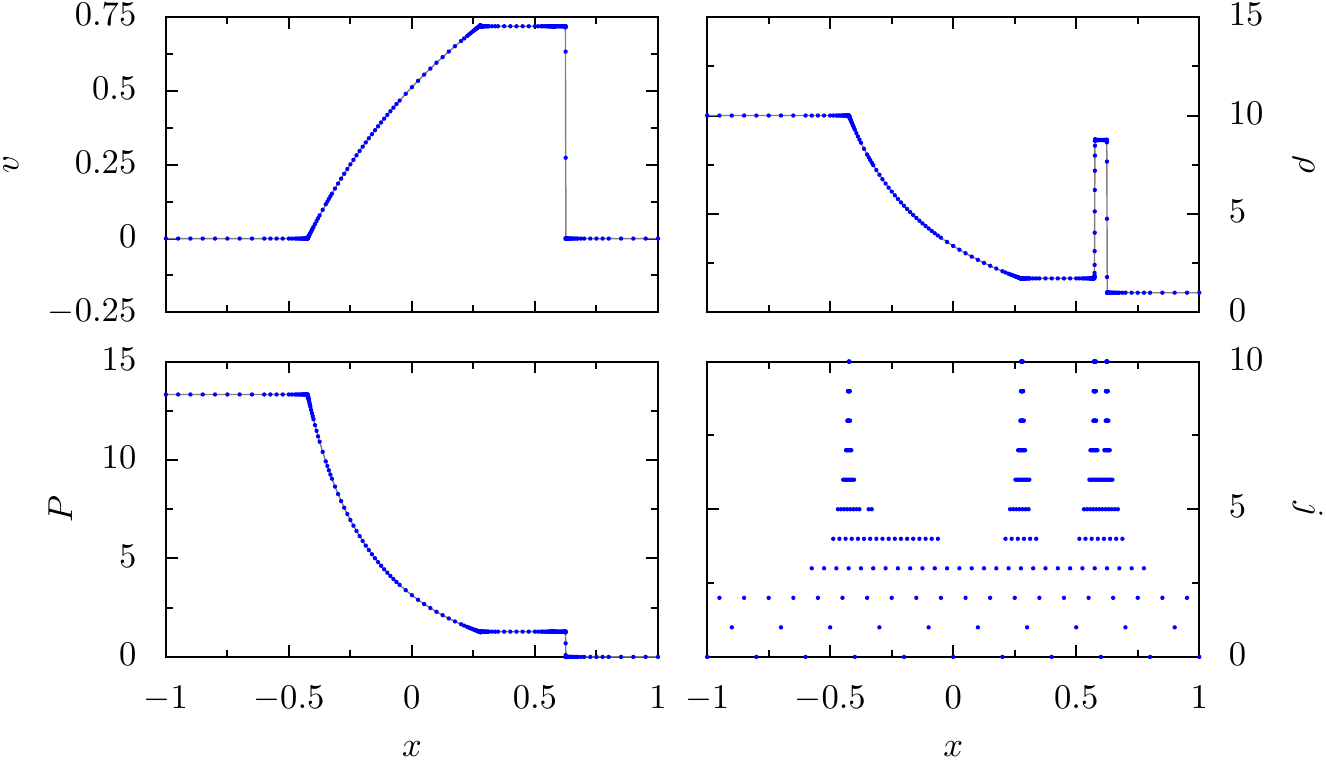}
  \caption{Comparison of the results from \oahu~to the exact solution for the
  Case I Riemann problem (see Table~\ref{tab:cases}). The solution (shown in
  gray) is at time $t = 0.8$. The positions of the wavelet grid points and their
  associated field values are shown in blue. This simulation was performed with
  $N = 10$, $j_{\rm max} = 10$ and $\epsilon = 10^{-5}$. The velocity, pressure and
  density all match the exact solution well. In the lower right panel the
  level of the point is plotted against the position of the point. The features
  in the profiles of $v$, $P$, and $\rho$ match the location of refinement in the grid.
  The resulting wavelet grid is not hand-tuned; instead, it adapts to the
  evolution of the fluid. At the time shown, only 324 out of a possible 10241
  grid points are occupied.}
  \label{fig:caseIcompare}
\end{figure*}

In each case, the results obtained with \oahu~match closely the exact solution.
Figure~\ref{fig:caseIcompare} shows the results for Case I at $t = 0.8$ with
$N = 10$, $j_{\rm max} = 10$ and $\epsilon = 10^{-5}$. The solution found by \oahu~
matches the exact solution extremely well. The final grid has adapted to the
features that form during the simulation. The final state shown has only
324 points out of a possible 10,241 points, giving this simulation a very
high effective resolution at a savings of over 96 percent.

The bottom right panel of Figure~\ref{fig:caseIcompare} gives the level of
each point in the grid. This figure is characteristic of the refinement of
the wavelet method. When the solution is smooth, there is very little
refinement, and where the solution exhibits sharp features, the refinement
proceeds to higher levels. For the very sharp features in this example, the
refinement proceeds to the highest level allowed for the particular simulation.
A discontinuity in the solution will generically refine as far as is allowed
by the simulation. There is no smooth approximation at any resolution to a
sharp transition.

\begin{figure}[htb]
  \includegraphics[width=3in]{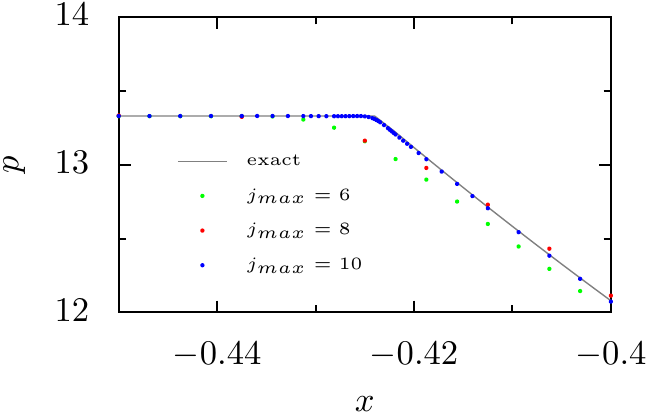}
  \includegraphics[width=3in]{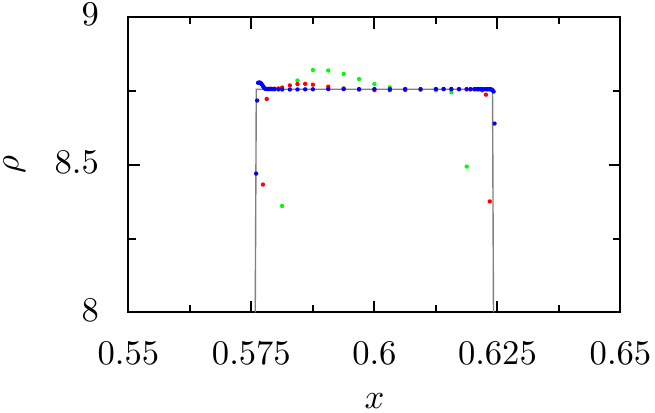}
  \caption{Detail of pressure at one end of the rarefaction wave (top)
  and the density in the shock (bottom) for Case I, with three different
  values of $j_{\rm max}$. As the maximum refinement level is increased, the
  wavelet solution matches the exact solution more closely. }
  \label{fig:caseIdetail}
\end{figure}

\begin{figure}[htb]
  \includegraphics[width=3in]{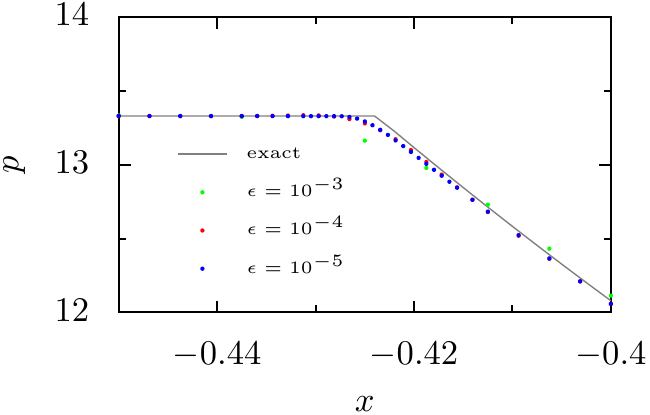}
  \caption{Detail of the pressure similar to the top panel of
  Figure~\ref{fig:caseIdetail}, but with fixed $j_{\rm max} = 8$ and a varying
  $\epsilon$. The refinement criterion has already pushed the refinement to
  the maximum level with $\epsilon = 10^{-4}$, and so there is little change
  as $\epsilon$ is decreased to $10^{-5}$.}
  \label{fig:caseIdetaileps}
\end{figure}

It is interesting to explore how the quality of the solution depends on the
maximum refinement level allowed and on the refinement criterion, $\epsilon$.
Figure~\ref{fig:caseIdetail} shows two close up views of features for
Case I run for three different values of $j_{\rm max}$. In each case, $N = 10$
and $\epsilon = 10^{-5}$. The overshoot in the density for the $j_{\rm max} = 10$
case is not an artifact of the adaptive scheme we employ; when run at an
equivalent resolution in unigrid ($N = 5120$ and $j_{\rm max} = 1$), the same
overshoot is present. A similar close up is shown in
Figure~\ref{fig:caseIdetaileps} giving a comparison of a set of runs with
$N = 10$, $j_{\rm max} = 8$ and various values of $\epsilon$. As $\epsilon$
decreases, the solution matches more closely the exact solution. However, there
is little difference between $\epsilon = 10^{-4}$ and $10^{-5}$. This
demonstrates the interplay between $j_{\rm max}$ and $\epsilon$. In this case, the
sharpened refinement criterion would drive more refinement near this feature,
but the maximum refinement level has already been reached.

Table~\ref{tab:errormeasures} gives details of an error measure for each
simulation:

\begin{equation}
L_2(f) = \frac{1}{N_{\rm occ}} \sqrt{ \sum_{j,k} (f(x^{j,k}) - f_{ex}(x^{j,k}))^2},
\end{equation}

\noindent where $N_{\rm occ}$ gives the number of occupied points, $j$ and $k$
index each point, $f(x)$ is the computed solution and $f_{ex}(x)$ is the
exact solution. Increasing the maximum refinement level at the same
$\epsilon$ tended to decrease the overall error. Similarly, as $\epsilon$ is
decreased, the error decreases. Though, as indicated in
Figure~\ref{fig:caseIdetaileps}, the refinement is reaching its allowed
maximum and so
the additional refinement that would be generated by the smaller $\epsilon$ is
not realized, leading only to modest accuracy gains. The much smaller errors
for Case III are not surprising as this test contains two rarefaction waves
and though there are sharp features in the exact solution, there are no
discontinuities.

\begin{table}[htb]
  \centering
  \begin{tabular}{|l|ccccccc|}
    \hline
     Case & $j_{\rm max}$ & $\epsilon$ & $N_{\rm occ}$ & $N_{\rm grid}$ & $L_2(v)$ & $L_2(\rho)$ & $L_2(P)$ \\
          &           &            &           &            & [$10^{-3}$] & [$10^{-3}$] & [$10^{-3}$] \\
    \hline
     I   & 6  & $10^{-5}$ & 201  & 641   & 2.88 & 29.1 & 3.41 \\
     I   & 8  & $10^{-5}$ & 261  & 2561  & 2.05 & 22.2 & 1.86 \\
     I   & 10 & $10^{-5}$ & 324  & 10241 & 0.89 & 21.1 & 1.87 \\
     I   & 8  & $10^{-3}$ & 145  & 2561  & 3.63 & 40.1 & 3.29 \\
     I   & 8  & $10^{-4}$ & 215  & 2561  & 2.50 & 26.9 & 2.19 \\
    \hline
     II  & 10 & $10^{-5}$ & 320  & 10241 & 2.16 & 21.3 & 2.19 \\
     III & 10 & $10^{-5}$ & 422  & 10241 & 0.009 & 0.003 & 0.002 \\
     IV  & 10 & $10^{-5}$ & 2759 & 10241 & 0.298 & 0.868 & 219 \\
    \hline
  \end{tabular}
  \caption{Statistics for the Riemann problems presented in this work. $N_{\rm occ}$
  gives the number of occupied grid points, while $N_{\rm grid}$ gives the maximum
  number of available points. The last three columns give the $L_2$-norms of
  the error of the velocity, density and pressure, as described
  in the text. Note that the MP5 reconstruction failed in Case IV, and so
  results using PPM reconstruction are given instead.}
  \label{tab:errormeasures}
\end{table}

\oahu~also monitors the conservation of the evolved variables during a
simulation. For these test cases, the conservation
is as good as the specified $\epsilon$. This is to be expected as the
representation keeps details only if those details are larger than $\epsilon$.
In each of the cases above, the relative drift in the conserved quantities
is on the order of $\epsilon$ for that case.

\section{Relativistic Kelvin-Helmholtz Instability}
\label{sec:kelhel}

We applied \oahu~to the relativistic Kelvin-Helmholtz instability in two
dimensions. For comparison, we have used identical initial conditions as
in~\cite{Radice2012}. The computational domain is taken to be a periodic
box from $x = -0.5$ to $x = 0.5$, and from $y = -1$ to $y = 1$. The shear
is introduced via a counterpropagating flow in the $x$ direction

\begin{equation}
v^x(y) = \begin{cases}V_s \tanh{\left[(y - 0.5)/a\right]}, & y > 0\\
                      -V_s \tanh{\left[(y + 0.5)/a\right]}, & y \le 0.
                    \end{cases}
\end{equation}

\noindent Here $a = 0.01$ is the thickness of the shear layer
and $V_s = 0.5$. A small perturbation of the velocity transverse to the shear
layer seeds the instability:

\begin{equation}
v^y(x,y) = \begin{cases}
    A_0 V_s \sin(2 \pi x) \exp\left[-(y - 0.5)^2/\sigma\right], & y > 0\\
    -A_0 V_s \sin(2 \pi x) \exp\left[-(y + 0.5)^2/\sigma\right], & y \le 0,
  \end{cases}
\end{equation}

\noindent where $A_0 = 0.1$, and $\sigma = 0.1$. For this test, $\Gamma = 4/3$
and the pressure is initially constant, $P = 1$. The density is given by a
profile similar to $v^x$ superposed on a constant as follows

\begin{equation}
\rho(y) = \begin{cases}
    \rho_0 + \rho_1 \tanh\left[(y - 0.5)/a\right], & y > 0\\
    \rho_0 - \rho_1 \tanh\left[(y + 0.5)/a\right], & y \le 0,
  \end{cases}
\end{equation}

\noindent where $\rho_0 = 0.505$ and $\rho_1 = 0.495$.

\begin{figure}
 \includegraphics[width=3in]{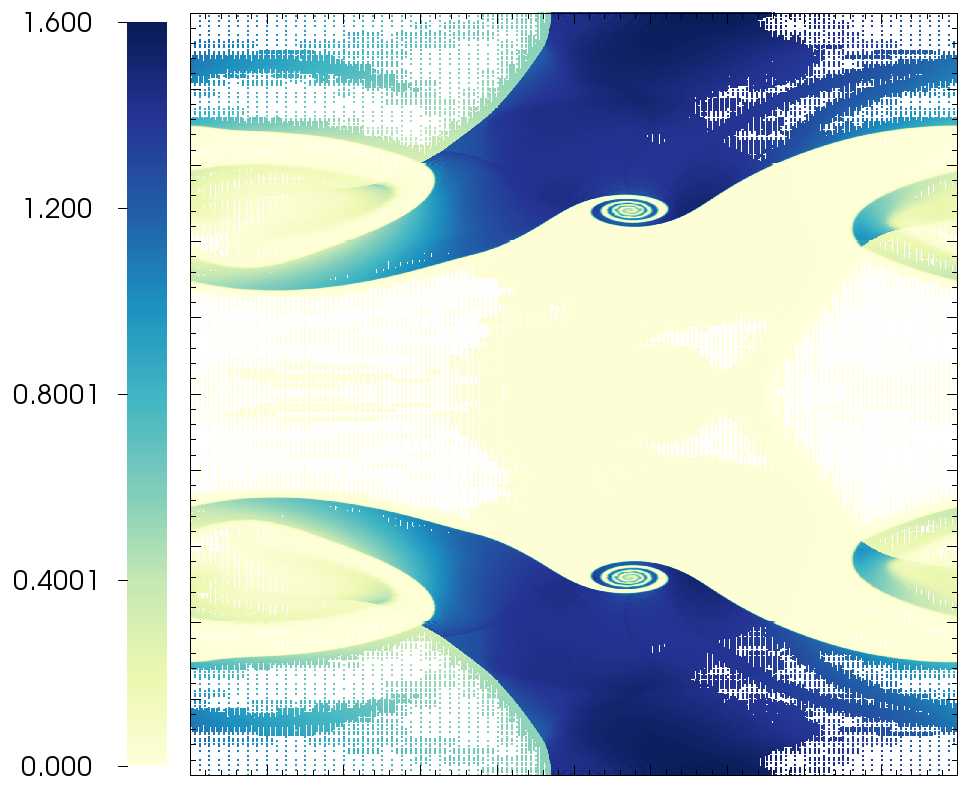}
 \caption{The relativistic Kelvin-Helmholtz instability at time $t = 3$.
  The points are colored by their density. This simulation used
  $(N_x, N_y) = (40, 80)$, $j_{\rm max} = 6$ and $\epsilon = 10^{-4}$.
Compare with~\cite{Radice2012}.}
 \label{fig:khrho}
\end{figure}

Shown in Figure~\ref{fig:khrho} is the density of this system at time $t = 3$
for a run having a base grid size of $(N_x, N_y) = (40, 80)$, and a maximum
refinement level of $j_{\rm max} = 6$. Thus, the effective grid size is
$2560 \times 5120$. The refinement criterion was $\epsilon = 10^{-4}$.
The final state is consistent with the
results in~\cite{Radice2012}. The conservation of the fluid is consistent with
the chosen $\epsilon$: $D$ suffers a relative change of $6.13 \times 10^{-6}$,
and $\tau$ suffers a relative change of $1.04 \times 10^{-4}$.
Note the appearance of the secondary whirl along
the shear boundary. In accord with the results of~\cite{Radice2012}, we find
that the number and appearance of these secondary instabilities depend on
the maximum resolution employed in our simulation, supporting their finding
that these secondary instabilities are numerical artifacts.

\section{Rayleigh-Taylor instability in a relativistic outflow}
\label{sec:outflow}

One interesting problem in relativistic fluid dynamics is the interaction
of a relativistic blast wave of ejecta from a GRB explosion
with the surrounding ISM. As the shock expands adiabatically into the ISM,
it loses thermal energy and eventually decelerates.
This results in a double-shock system with a forward
shock (traveling into the ISM), a contact discontinuity, and a reverse
shock (moving into the ejecta). \citet{Levinson:2010wz} showed that
the contact discontinuity is unstable to the Rayleigh-Taylor instability.
The turbulence generated by this instability can amplify magnetic
fields and the emission from the thin shell of material behind the forward
shock. Duffell and MacFadyen have studied this system with numerical
simulations in a series of
papers~\citep{Duffell:2011bc,Duffell:2013ckh,Duffell:2014vsa},
finding that the Rayleigh-Taylor
instability can disrupt the forward shock for soft equations of state,
which might be typical of radiative systems~\citep{Duffell:2014vsa}.

Simulating the relativistic outflow of GRB ejecta
constitutes an especially challenging numerical test for an adaptive
relativistic fluid code. Relativistic effects compress the width of the
thin shell by the Lorentz factor squared, $\Delta r/r \approx 1/W^2$.
Capturing the Rayleigh-Taylor instability that forms at the
contact discontinuity within the shell thus requires very high resolution
within a thin shell that propagates outward with a velocity near the speed of
light. Duffell and MacFadyen succeeded in simulating this system using
an elegant moving mesh code, {\sc TESS}~\citep{Duffell:2011bc}.
The computational cells in {\sc TESS} are allowed to move with the fluid,
giving very high resolution in the shell and at the shocks.
As a final test, we repeat the decelerating shock test here with
\oahu~to demonstrate the adaptive capability of the wavelet approach
for relativistic hydrodynamics.

We use the initial data for the decelerating shock given
in \citet{Duffell:2011bc}.
The initial data are spherically symmetric, but this symmetry is broken by
the instability, so we perform the simulation in cylindrical coordinates.
We label these coordinates $\{s,z\}$, where $s$ is the cylindrical radius,
and the spherical radius $r$ is given by $r^2 = s^2 + z^2$.
The initial data for the spherical explosion are
\begin{equation}
\arraycolsep=1.4pt\def\arraystretch{2.0}
\rho = \left\{\begin{array}{ll}
          \rho_0(r_0/r_{\rm min})^{k_0} & r < r_{\rm min} \\
          \rho_0(r_0/r)^{k_0}              & r_{\rm  min} < r < r_0 \\
          \rho_0(r_0/r)^k                  & r_0 < r,
       \end{array}
       \right.
\end{equation}
where the different parameters are chosen to be
\begin{equation}
k_0 = 4, \quad r_0 = 0.1, \quad r_{\rm min} = 0.001.
\end{equation}
A spherical explosion into a medium with a power-law dependence,
$\rho \simeq r^{-k}$, will decelerate if $k < 3$. So we choose
$k=\{0,1,2\}$ for our runs below.
The pressure is given by
\begin{equation}
\arraycolsep=1.4pt\def\arraystretch{2.0}
P = \left\{
    \begin{array}{ll}
       e_0\rho/3  & \mbox{for } r < r_{\rm exp}\\
       10^{-6}\rho & \mbox{for } r > r_{\rm exp},
    \end{array}
    \right.
\end{equation}
where $r_{\rm exp} = 0.003$
and the constant $e_0$ is chosen such that the outgoing shock has a
Lorentz factor of $W \approx 10$. We set $e_0 = \{6, 4, 6\}$ for
$k=\{0, 1, 2\}$, respectively.

\begin{figure*}
\centering
\includegraphics[width=0.5\linewidth]{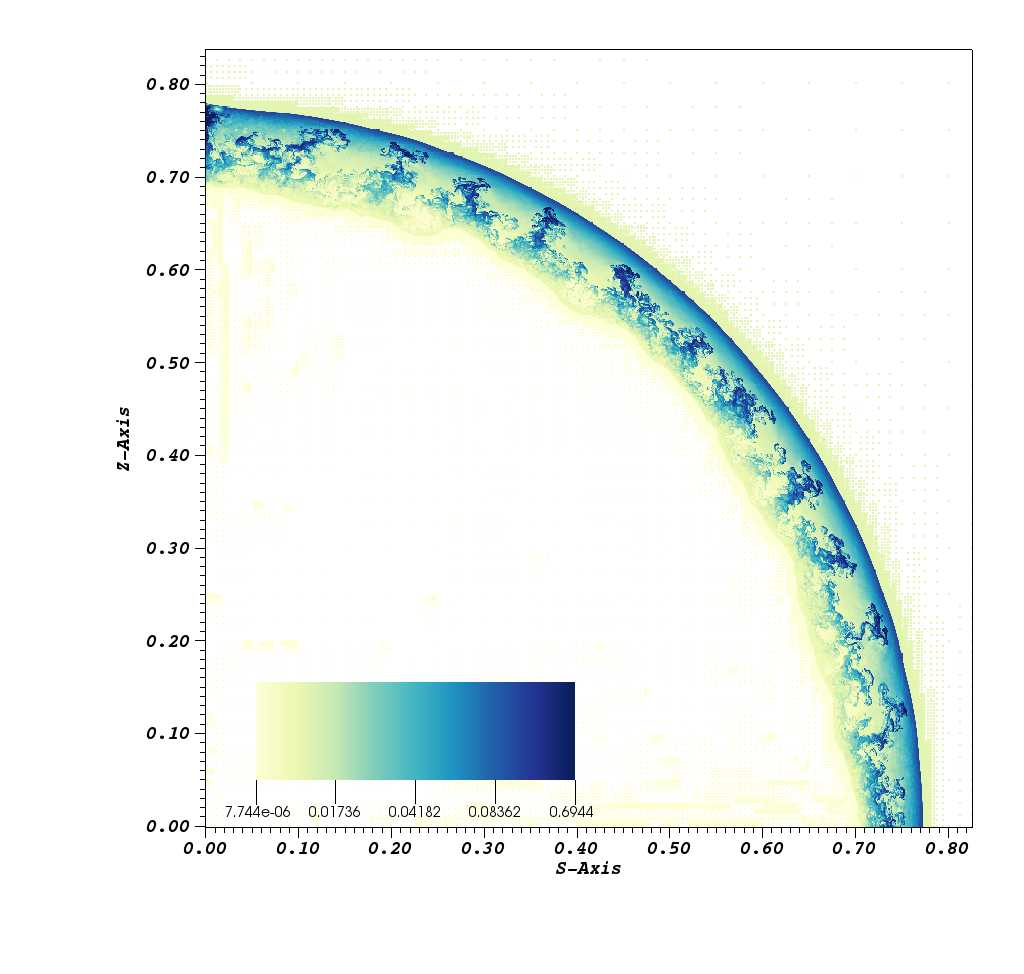}
\includegraphics[width=0.49\linewidth]{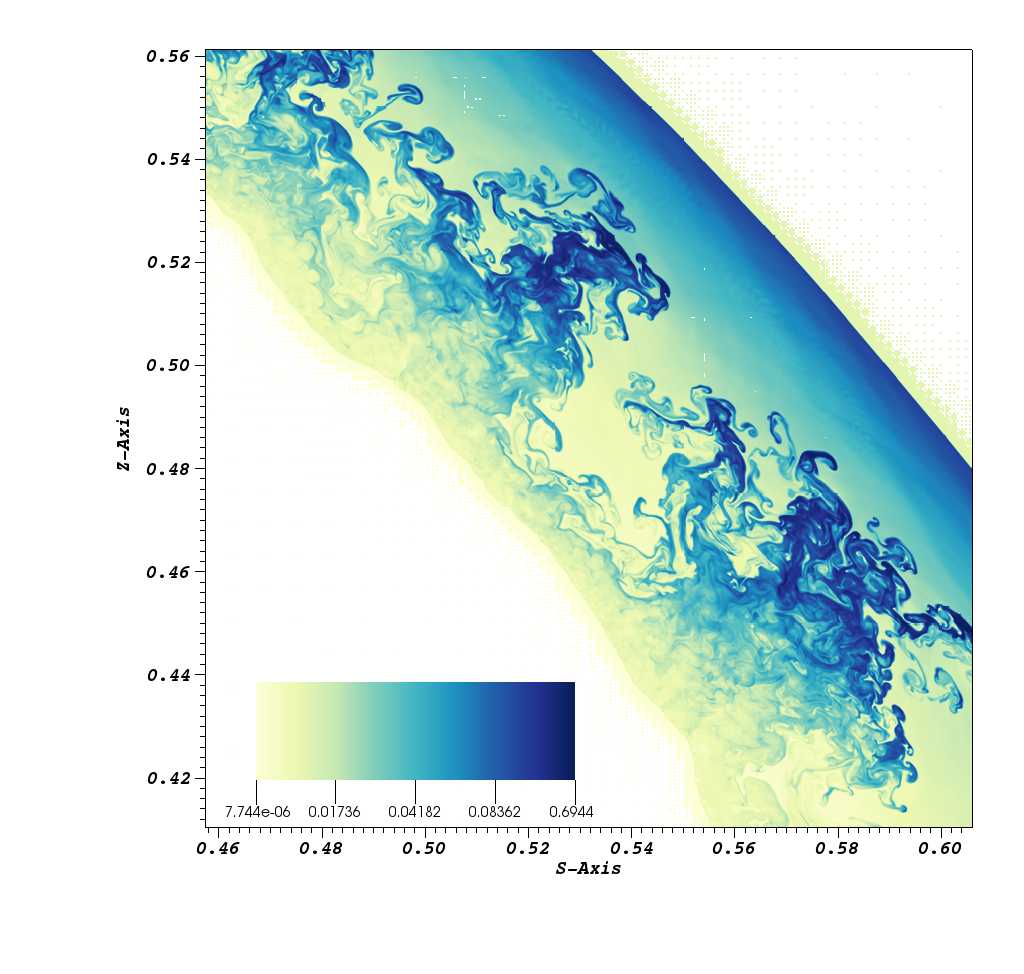}
\caption{The $k=0$ spherical explosion case with a decelerating
relativistic shock and a mass excess. After a coasting period, the
solution develops two shocks separated by a contact discontinuity unstable
to the Rayleigh-Taylor instability.  Results are shown at $t=0.8$ for
$j_{\rm max} = 10$ with $\epsilon=10^{-4}$. This figure also demonstrates the
adaptivity of the method; outside the blast wave there is little refinement,
while inside the blast, the small scale features drive high levels of
refinement, leading to a well-resolved Rayleigh-Taylor instability.}
\label{fig:k0}
\end{figure*}

\begin{figure*}
\includegraphics[width=0.5\linewidth]{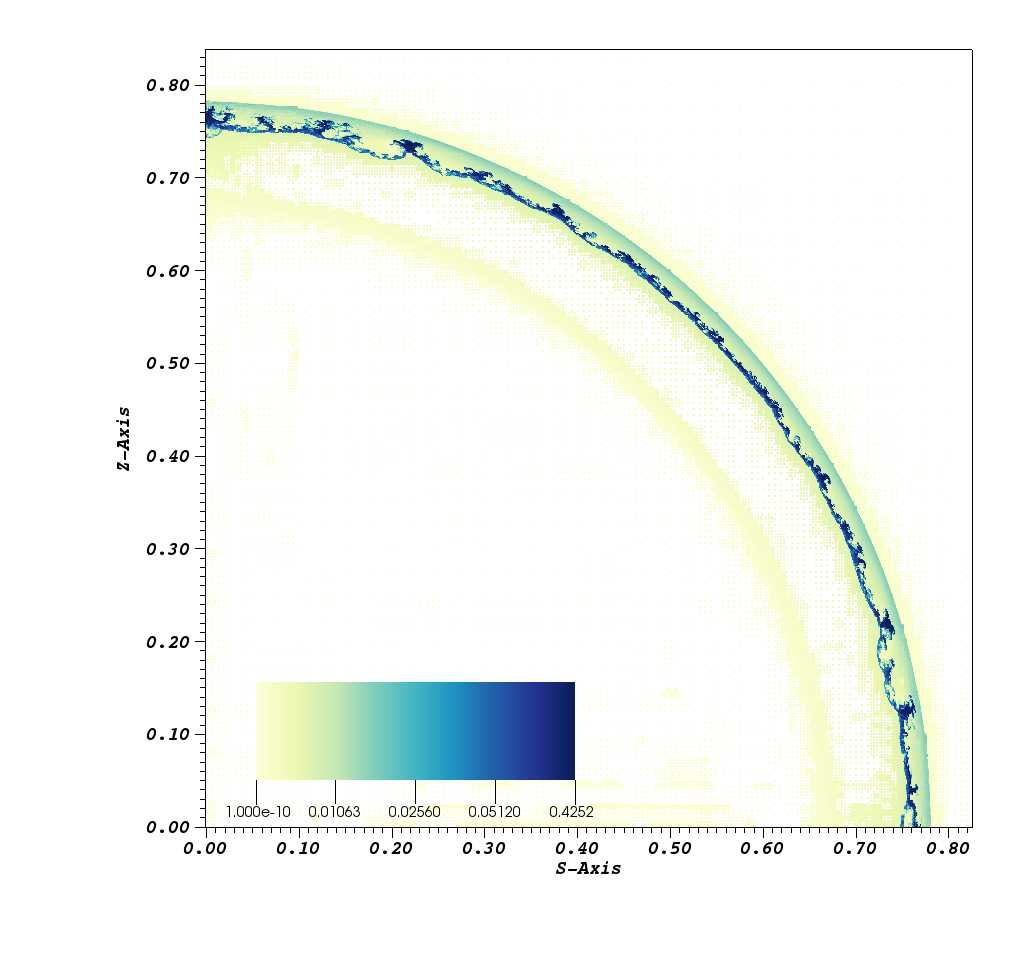}
\includegraphics[width=0.5\linewidth]{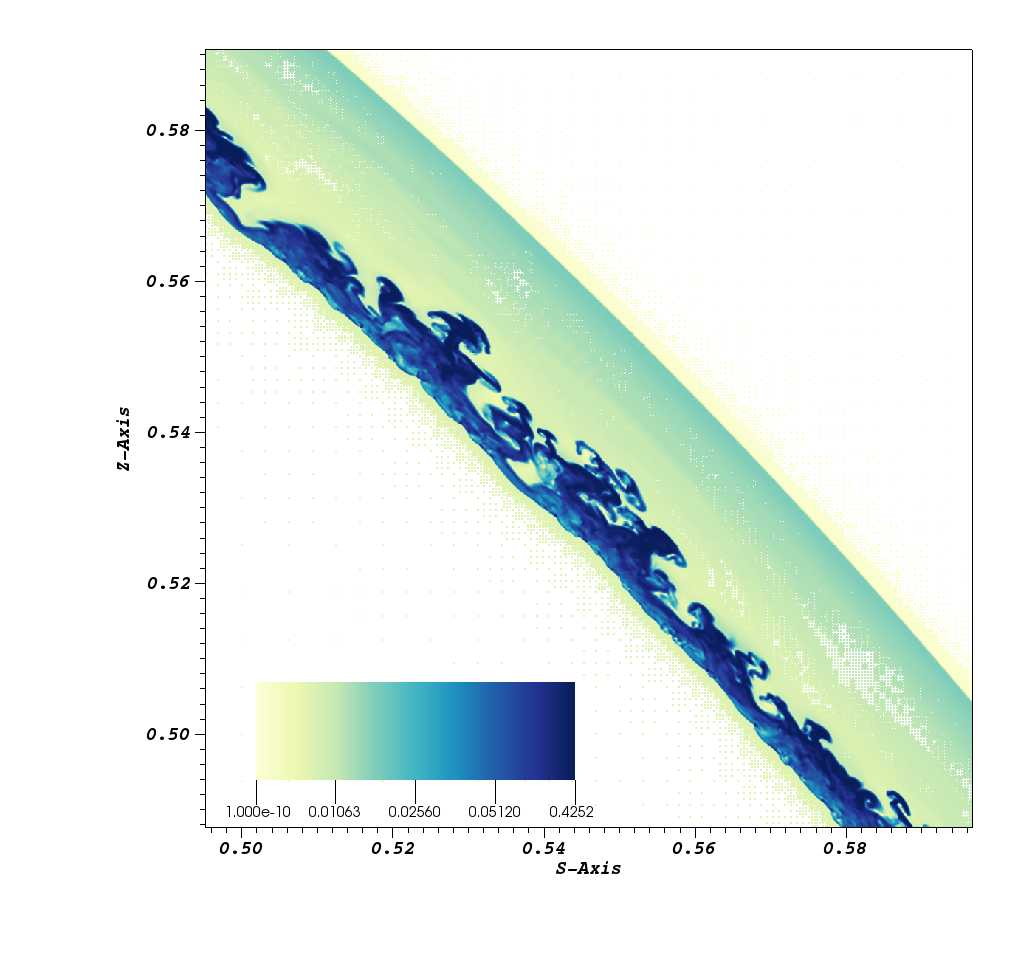}
\caption{The $k=1$ spherical explosion case with a decelerating
relativistic shock and a mass excess.  Results are shown at $t=0.8$ for
$j_{\rm max} = 10$ with $\epsilon=10^{-4}$.}
\label{fig:k1}
\end{figure*}

\begin{figure*}
\includegraphics[width=0.5\linewidth]{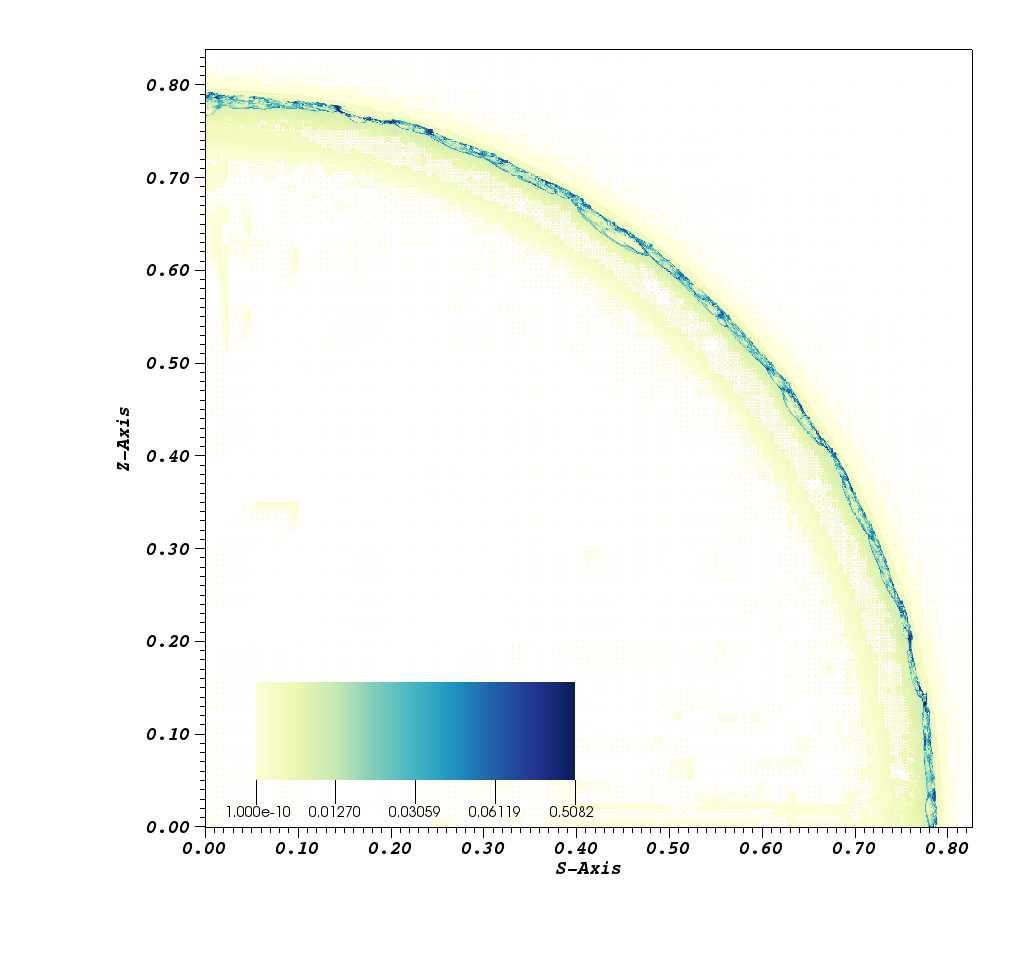}
\includegraphics[width=0.5\linewidth]{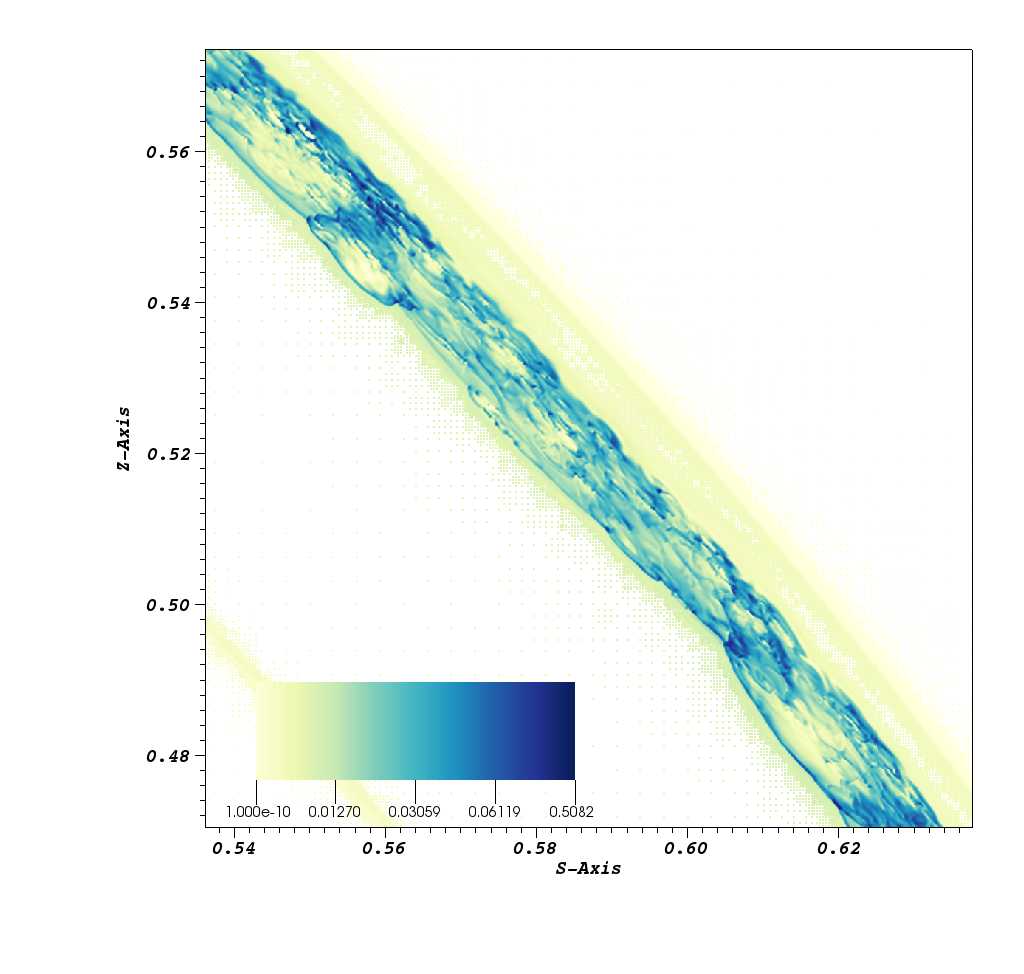}
\caption{The $k=2$ spherical explosion case with a decelerating
relativistic shock and a mass excess.  Results are shown at $t=0.8$ for
$j_{\rm max} = 10$ with $\epsilon=10^{-4}$.}
\label{fig:k2}
\end{figure*}

Three different cases $k=\{0, 1, 2\}$ are simulated to $t=0.8$, and the
solutions at this time are shown in
Figures~\ref{fig:k0},~\ref{fig:k1},~\ref{fig:k2}.  In each case, ten levels
of refinement were used with a refinement criterion value of
$\epsilon=10^{-4}$.  The Lorentz factor for the shock waves is $\sim 12$.
All simulations
were evolved in their entirety in 2-D from the initial conditions
using the wavelet method described in this work.  The results are consistent
with those obtained with the TESS code.
As $k$ is increased, the width of the blast wave decreases. For $k=0$ the
RT instability is well resolved. For $k=1$, the instability is evident, though
with less detail than $k=0$. In the $k=2$ case, the width of the blast wave is
too narrow to properly resolve the substructures in the instability. That there
is an instability is apparent, but it lacks the definition of even the $k=1$
case. These features could be resolved by increasing $j_{\rm max}$ for the $k=2$
simulation. The conservation of all conserved variables was externally monitored
throughout the simulation.  Variation in the conservation of $D$
amounted to less than 0.001\% while variation in the conservation of $\tau$
was somewhat larger at 0.8\%, visibly due to boundary effects along the $z$-axis.


\section{Summary}
\label{sec:summary}

Motivated by the need to efficiently resolve the many emerging
localized features and instabilities present in astrophysics simulations
such as the merger of two neutron stars, this work has presented a
wavelet based approach for solving the relativistic hydrodynamic equations.
The resulting implementation of this approach, \oahu, has
reproduced a number of results in relativistic hydrodynamics, including the
one dimensional shock tube tests of~\cite{LoraClavijo2013}, the relativistic
Kelvin-Helmholtz instability~\citep{Beckwith2011,Radice2012}, and
the Rayleigh-Taylor instability resulting from a gamma ray burst outflow
model~\citep{Duffell:2011bc}.  Unlike adaptive mesh refinement based on nested
boxes, the unstructured dyadic grid of collocation points in the wavelet
approach
conforms to
highly localized solution features without creating the box-shaped numerical
artifacts typically present in nested box adaptive mesh refinement simulations.
Further, the approach presented here demonstrates the efficiency and utility
of using the coefficients from the wavelet transformation to drive refinement
without requiring problem specific \emph{a priori} refinement criteria.

The wavelet method described here can be directly applied to
the equations of relativistic magnetohydrodynamics (MHD) which describe
a plasma of relativistic particles in the limit of infinite conductivity.
Investigating MHD with wavelets will be part of future work.

For use in the merger simulations of astrophysical
compact objects such as neutron stars and black holes,
the wavelet based relativistic hydrodynamics kernel must be integrated with
a kernel solving the Einstein equations for gravity.  The wavelet method
presented in this work has been designed expressly for this purpose and
is a crucial feature for its use in astrophysics.  The results from
fully dynamic gravitational and hydrodynamics simulations as well as the
method for integrating the hydrodynamics and gravitation computational kernels
with the wavelet approach will be reported in future work.

Although not a focus of this work, the parallel implementation
of the wavelet simulation framework presented here
deviates from conventional practice in combining multi-threading with
a form of message-driven computation sometimes referred to as
asynchronous multi-tasking.  The
scalable asynchronous multi-tasking aspects of this work
will be addressed in future work.

\section{Acknowledgments}
It is a pleasure to thank our long-term collaborators Luis Lehner,
Steven L. Liebling, and Carlos Palenzuela, with whom we have had many
discussions on adaptive methods for relativistic hydrodynamics.  We
acknowledge Thomas Sterling with whom we have had many discussions on
the parallel implementation of \oahu.  We also acknowledge many helpful
discussions on wavelets with Temistocle Grenga and Samuel Paolucci.
This material is based upon work supported by the Department of
Energy, National Nuclear Security Administration,
under Award Number DE-NA0002377,
the Department of Energy under Award Number DE-SC0008809, the
National Science Foundation under Award Number PHY-1308727,
and NASA under Award Number BL-4363100.

\bibliographystyle{plainnat}
\bibliography{paper}

\end{document}